\documentstyle[aps,graphicx,prl]{revtex} 
\tighten 
\begin{document} 
\draft 
\twocolumn[\hsize\textwidth\columnwidth\hsize\csname 
@twocolumnfalse\endcsname 
\title{The Mesoscopic Chiral Metal--Insulator Transition}  
\author{S. Kettemann$^{(1)}$, B. 
  Kramer$^{(1)}$, T.  Ohtsuki$^{(1,2)}$ }  
\address{ $^{1}$I. Institut f\"ur 
  Theoretische Physik, Universit\" at Hamburg, Jungiusstra\ss{}e 9, 20355 
  Hamburg, Germany\\ 
 $^{2}$Department of Physics, Sophia 
  University, Kioi-Choi 7-1, Chiyoda-Ku, Tokyo 102-8554, Japan} 
\maketitle 
\begin{abstract} 
  Sharp localization transitions  of  chiral edge states 
 in disordered quantum wires,  subject to  strong magnetic field,  are 
  shown  to be 
  driven by crossovers from two- to one-dimensional localization of  bulk
  states.
 As a result, the two-terminal conductance 
  is found  to 
 exhibit at zero temperature discontinuous transitions between {\it exactly}
  integer
  plateau values and zero, reminiscent of first order phase transitions.
 We discuss the corresponding phase diagram.  The 
  spin of the electrons is shown to result in  a   multitude of phases, when 
  the spin degeneracy is lifted by the Zeeman energy. 
 The width of  conductance plateaus is  found to depend sensitively
 on the spin flip rate $1/\tau_s$. 
\end{abstract} 
\vskip2pc] 
 
The high precision of the quantization of the Hall conductance of a two
dimensional (2D) electron system (2DES) in  a strong  magnetic field
\cite{klitzing} is known to be due to the binding of electrons to localized
states in the bulk of the 2DES. Thereby,  a change of electron density does not
 change the  Hall conductance \cite{laughlin,aoki,halperin}.
The localization length in  tails of  Landau bands is very small,
of the order of the cyclotron length
$l_{\rm cyc} = v_F/\omega_B = \sqrt{2 n+ 1} l_B$. 
  It increases towards the centers of the Landau
bands, $E_{n0} = \hbar \omega_B (n +1/2)$ ($n=0,1,2,...$), with $\omega_B = e
B/m^{*}$ the cyclotron frequency ($e$ elementary charge, $m^{*}$ effective
mass),  $v_F$ the Fermi velocity, 
and $l_B^2 = \hbar/e B$ defines the magnetic length.
  In an {\em infinite} 2DES in 
perpendicular magnetic field, the localization length at energy $E$ diverges
as $ \xi \sim |E -E_{n0}|^{-\nu}$, reminiscent of 2nd order phase
transitions.  The critical exponent $\nu$ is found numerically for the lowest
two Landau bands, $n=0,1$, to be $\nu = 2.33\pm0.04$ for spin-split Landau
levels \cite{hucke,huckerev}.  Analytical \cite{raikh} and
experimental studies \cite{scalingexp} are consistent with this value. In a
{\em finite} 2DES, a region of extended states should exist in the centers of
 disorder broadened Landau-bands. These states extend beyond the system
size $L$. The width of these regions is given by $\Delta E = (l_{\rm
  cyc}/L)^{1/\nu} \Gamma$, where   $\Gamma = \hbar ( 2
\omega_B/\pi\tau)^{1/2}$ is the band width,
with elastic scattering time $\tau$. 
 
\begin{figure}
\begin{center}
\vspace{-4.5cm}
  \includegraphics[width=.45 \textwidth ]{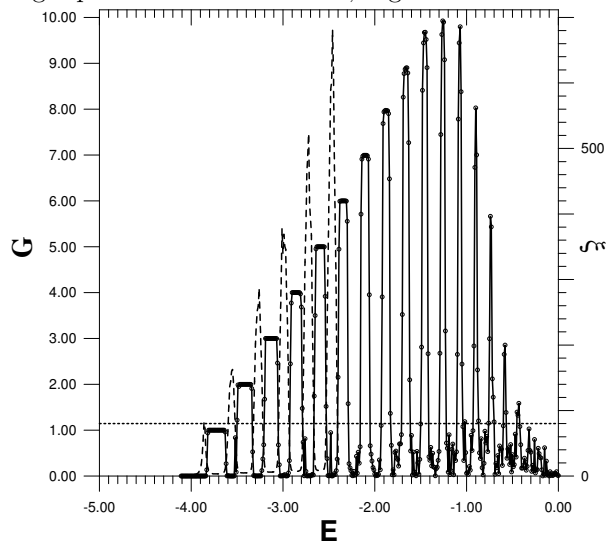}
\vspace{-0.cm}
\caption{ Solid curve, left axis: Two--terminal conductance
  of  a wire  (width $L_y = 80 a$,  length $L = 5000 a$, hard wall
  confinement)
 in units of $G_0= e^2/h$ as  function of
  energy $E$ in units of $t$ (hopping amplitude in the tight binding
  model). Disorder strength,  $W = .8 t$. There are  $x = .025$ 
 magnetic  flux quanta through an elementary
  cell $a^2$.
 Dashed curve, right axis: 
  bulk localization length $\xi$ in units of
   $a$ as  function of
   $E$, from   transfer matrix method for  wire of
  identical properties but periodic boundary conditions and $L_{max} = 100 000
  a$.  Straight dashed  line:   $L_y = 80 a$.
   }
 \label{loclength}
\end{center}
\vspace{-.5cm}
\end{figure}

In quantum Hall bars of finite width, there exist in addition 
edge states with energies lifted by the confinement potential above the
energies of  centers of  bulk Landau bands, $E_{n0}$ \cite{halperin}.
Previously, there has been a considerable interest in the study of
mesoscopically narrow quantum Hall bars\cite{haug}, with  emphasis on conductance
fluctuations\cite{timp,ando94}, edge state mixing
\cite{edgemixing,shkledge,ohtsuki,mani},  breakdown of the quantum Hall
effect\cite{breakdown}, and  quenching of the Hall effect due to classical
commensurability effects \cite{roukes}.
It is known  that in the presence of white noise disorder
the edge states do mix with the bulk states when the Fermi energy is moved
into the center of a Landau band. It had been suggested that this might result
in localization of edge states \cite{ohtsuki,mani}.
In this paper, we show  that this is
indeed the case. In particular, at zero temperature the
two-terminal conductance  of a quantum wire in a
magnetic field exhibit for uncorrelated disorder and hard wall confinement
discontinuous transitions between integer plateau values and zero,  Fig \ref{loclength}.

The localization length $\xi$ in a 2DES with broken time reversal symmetry is
expected from  scaling theory\cite{ab,weg,hikami,elk} and
numerical scaling studies \cite{mk81,transfer} to be 
\begin{equation} \label{2db} 
\xi_{\rm 2D}  = l_0\, e^{\pi^2 g^2}. 
\end{equation} 
 depending exponentially on 
 $g$,  the 2D conductance parameter per spin channel.  
 $l_0$ is the short distance cutoff, which
is the elastic mean free path $l = 2 g(B=0)/k_{{\rm F}}$ ($k_{\rm F}$ Fermi
wave number) at moderate magnetic fields, $b\equiv\omega_B \tau < 1$.
 For stronger magnetic fields, $b>1$,   $l_0$ crosses
over to the cyclotron length $l_{\rm cyc}$.
   $g$ exhibits Shubnikov-de-Haas oscillations as function
of  magnetic field for $b>1$. Maxima occur when the Fermi energy is in
the center of  Landau bands. Thus, the localization length 
increases strongly from  band tails to  band centers, even 
 when  the wire  width $L_y$  is too narrow to allow delocalisation 
of bulk states. For uncorrelated impurities, within self
consistent Born approximation \cite{ando},  the maxima 
 are given by $g(E = E_{n 0}) =  (2  n + 1)/\pi = g_n
$.  Thus, $ \xi_{\rm 2D} (E_{n 0}) = l_{\rm cyc} \exp{( \pi^2 g_n^2 )} $ are macroscopically
large in  centers of higher Landau bands, $n > 1 $
\cite{huckerev,levitation}.  However, when the width of the system $L_y$ is
smaller than  $\xi_{\rm 2D}$, localization is expected to
behave quasi-1D. In other words,  electrons in  centers of  Landau
bands can diffuse between the edges of the system, but are localized parallel
to the  edges if $L_y < \xi_{\rm 2D}$. 
 The quasi-1D localization length is known to
depend only linearly on $g$. In a magnetic field, when
time reversal symmetry is broken, it is \cite{ef,larkin,dorokhov}
\begin{equation} \label{1db} 
\xi_{\rm 1D} = 2 g (B)  L_y.   
\end{equation} 
There is a crossover from 2D to 1D
localization as the Fermi energy is moved from  tails to  centers of 
Landau bands. 
 Performing  a    renormalization  of the wire conductance, 
 one obtains for the  localization length \cite{crossover},
\begin{equation}  
\label{lcb} 
\xi^2  = 4  L_y^2 g^2 - \frac{2 L_y^2 }{ \pi^2} 
 \ln  \left[ \frac{1+ (L_y/ l_0 )^2}{ 1 + ( L_y/  \xi )^2} \right].   
\end{equation}  
 Its solution shows   a crossover between the quasi-1-D--
 and the 2D--behaviour,
 Eq.  (\ref{1db}) and  
 Eq. (\ref{2db}), respectively.

\begin{figure}[htbp]
\begin{center} 
\vspace{-1cm} 
\includegraphics[width=.4 \textwidth]{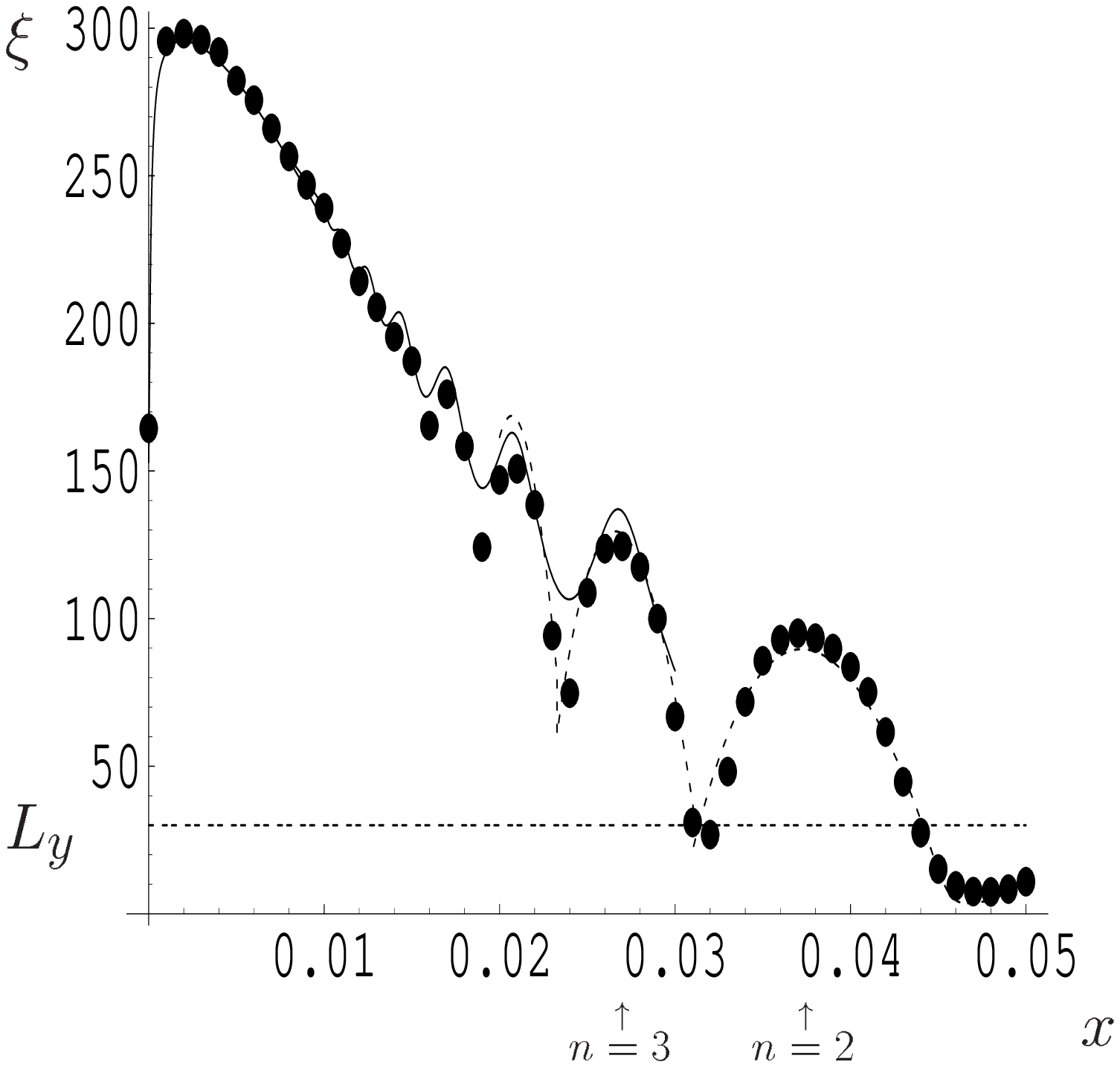} 
\vspace{-4.3cm} 
\caption[]{  The 
  localization length, $\xi$, as obtained by inserting
  $g(B)$ in 2nd order Born approximation in Eq.~(\ref{lcb}), including a
  summation over all Landau levels (full line). Broken line:
 $\xi$, with   $g(B)$ in 
 SCBA,  neglecting overlap between  Landau bands, Eq.
  (\ref{scba}).  The result of 
  transfer matrix calculations 
  is plotted  (dot size: numerical error $\approx 1\%$) for periodic boundary
  conditions \cite{kwaz93} as function of magnetic flux
 through  unit cell 
  $x = a^2/2 \pi l_B^2$. Dimensionless conductance
parameter  per spin channel  is $g(B=0) = 5.1$.  At
  weak magnetic fields  doubling of $\xi$ due to breaking of 
  time reversal symmetry is seen, in good agreement  with 
 the  analytical  crossover   formula (full line) \cite{crossover,prbr}. 
 Horizontal dashed  line: $L_y= 30 a$. }
 \label{lcbosc} 
\vspace{-.5cm} 
\end{center} 
\end{figure}

The conductance  per spin channel, $g(b) =
\sigma_{xx}(B)/\sigma_0$, is given by the Drude formula $ g(b) =
g_0/(1+b^2)$, ($g_0 = E \tau/\hbar$, $b=\omega_B \tau$) for weak magnetic field, $b
<1$.  For $b>1$, when the cyclotron length $ l_{\rm cyc}$ is smaller than the
mean free path $l$, disregarding the overlap between  Landau bands, $g$ is
obtained in SCBA \cite{ando},
\begin{equation} \label{scba} 
g(B)  = \frac{1}{\pi}  \left(2 n+1 \right) \left( 1 - \frac{(E_{\rm F}- 
 E_n)^2}{\Gamma^2}\right),  
\end{equation} 
for $ \mid E - E_n \mid < \Gamma$.  One obtains the localization length for $b>1$ and
$\mid \epsilon/b -n -1/2 \mid < 1$ by inserting $g$,  Eq.~(\ref{scba}) into Eq.~(\ref{lcb}).
 It oscillates between maximal values in
 centers of  Landau bands, and minimal values in  band tails  (Fig.~\ref{lcbosc}).
  For $n>1$, one finds in  band centers,
\begin{equation} \label{middle} 
\xi_n =  \frac{2}{\pi} \left( 2 n + 1 \right) L_y   
\left[ 1- \frac{\ln 
\sqrt{ 1+ \left(L_y/l_{\rm cyc} \right)^2}}{(n+1/2)^2} \right]^{1/2}.    
\end{equation}

In the center of the lowest  Landau band ($n=0$), Eq.~(\ref{1db})
gives a value $\xi_0 ( B) \approx (2/\pi) L_y$, smaller
than $L_y$. There, the localization is 2D and the topological term
\cite{pruiskenrev} is  effective, leading to  criticality, and  diverging
localization lengths.  In a wire of finite width, $L_y$ the localization length $\xi$
saturates to the critical value $\xi_{\rm crit} \approx 1.2 L_y$
\cite{hucke,huckerev} larger than    $L_y$. Comparison with
$\xi_{n}$, Eq. (\ref{middle}), shows  that the  noncritical  quasi-1D localization
length exceeds   $\xi_{\rm crit}$
in all but the lowest Landau bands.
If all states in  centers of  higher Landau bands are localized along
the wire, the question arises if  there exist extended
states in the quantum Hall wire at all.  
 Consider  an annulus, with a
circumference larger than the localization length in the center of a Landau
band.  When a magnetic flux pierces  the annulus, localized states are
unaffected. Guiding centers of states which extend around the
annulus do shift in position and energy, however\cite{halperin}. 
 In a confined wire there exist chiral edge states
\index{chiral edge states}, which  extend around the annulus.  A magnetic
flux change  moves these states  down and up  inner and outer edges,
respectively. As shown above,  in the middle of the Landau band, the  electrons can
diffuse freely from edge to edge, but are localized along the annulus
 with $\xi > L_y$.  As a
consequence, when adiabatically changing the magnetic flux, an ``edge state'' 
has to  move from the inner to the outer edge, since it cannot enter the band of
localized states. 
 This fact has  been interpreted as  proof for the existence of an
extended bulk state, extending around the annulus and between  edges, at the
energy $ E_{m}$, with $\xi (E_m) = L_y$ \cite{halperin}.
 In the following, we   show that  rather 
  a transition from 
  extended chiral edge states to localised states occurs at these energies, $E_m$. 
 
Using the transfer matrix method \cite{transfer}, we have calculated the
localization length as function of energy $E$ in a tight binding model of a
disordered quantum wire in  perpendicular magnetic field, with periodic
boundary conditions, Fig. 1 (dashed curve, right axis).
  Indeed, its
maxima are seen to increase linearly with energy $E$ in
  agreement with  Eq. (\ref{middle}).  The transfer
matrix result for the   2-terminal conductance $G$ \cite{pmr92}
is shown in Fig. 1 (solid curve, left axis) for a wire of identical
properties, but  hard wall boundary
conditions and finite length $L$.

We verify that the condition $\xi(E_{m,p}) = L_y$, yields 
the energies $E_{m,p}$, $p = \pm$, at which $m$ edge states mix, and
transitions from the quasi-1D chiral metal to an insulator occur, as signaled by sharp 
jumps of $G$ in Fig. 1. Here, $m=n$
when this energy is above the bulk energy of the $n$-th Landau band,
$p=+$, and $m= n-1$ when it is below it, $p=-$.
  For $\xi  < L_y$,  backscattering between edges is exponentially 
suppressed and the localisation length of edge states 
 increases exponentially as $\xi_{\rm edge} = \xi \exp (L_y/\xi)$.  
  These results are  summarised in the phase diagram, Fig. 3 a, 
 where the value of  $G$, in units of $e^2/h$, 
 is given  as function of wire width $L_y$ and energy $E$ in units of $\hbar \omega_B$. 
 We find that  $G= m$, where $m$ is the number of extended edge states. 
 When  $L_y \le l_{cyc} \propto \sqrt{2 n+1}$,  all edge states are localised, 
  and all conductance plateaus collapse, $G=0$, as seen in Fig. 1, close to the middle of 
  the band, $E=0$. 

 Taking into account the spin, the  Zeeman splitting $E_n^+-E_n^- = g_Z \mu_B B$,
 lifts the spin degeneracy,  where  $g_Z$ is the material dependent Zeeman-g-factor, 
 and $\mu_B$  Bohr's magneton. 
Without spin flip scattering  edge states of different spins are not mixed.  
 The 
 phase diagram,  Fig. 3 b, is then a  
   superposition of   two   phase diagrams, Fig. 3 a, for each spin.
  There are phases where the conductance is
   equal to  the total number of edge channels, $G=m^+ + m^-$,
 when the bulk localisation length of electrons  $\xi^{\sigma}$,
  does not   exceed the wire width,  $\xi^{\sigma} < L_y$, for either spin $\sigma =\pm$.
 When,  $\xi^{\sigma} > L_y$ for 
 one spin $\sigma$ only,  the conductance is carried by the number of edge states 
 with  opposite spin, $G=m^{-\sigma}$, only.
 If $\xi^{\sigma} > L_y$ for both $\sigma =\pm$,
  the conductance vanishes,   $G=0$.
  With  strong  spin flip scattering,  
 all edge state mix with  bulk states for   $\xi^{\sigma} > L_y$, yielding $G=0$.  
 There are  conductance plateaus
  equal to the total number of edge states,
 $G=m^+ + m^-$, only if 
  $\xi^{\sigma} < L_y$ is fullfilled for both spins,   $\sigma =\pm$ (Fig. 3 c).
 Thus,  both the  sequence  
  and   width of  conductance plateaus are  
  sensitive measures of  the spin flip scattering rate $1/\tau_s$,  due to 
 electron-electron interaction, spin-orbit interaction, 
 or scattering from  nuclear spins\cite{komi}. 
 Furthermore,   the  enhancement of the
 $g_Z$--factor above its bulk value due to the exchange interaction 
   depends on the  Fermi energy\cite{fang}.
 Accordingly, the 
 width of the conductance plateaus  changes with  energy. 
 In a real sample there exists a slowly
varying potential disorder which can stabilize  edge states against
mixing with bulk states  \cite{ohtsuki}. When the
confinement potential is varying slowly on the magnetic length scale $l_B$,
the   energies $E_{m,p}$ are expected to split into $m$ energies,
 at which  edge
states mix one by one with  bulk states,  accompanied by steps of  heights $1$ in 
 conductance $G$. Both potentials can be 
 effectively modified by the Coulomb interaction as obtained
  by a selfconsistent solution of the Poisson equation. 
 \begin{figure}[]
\begin{center} 
\vspace{-2.6cm} 
\hspace{-1cm}
\includegraphics[width=.8 \textwidth]{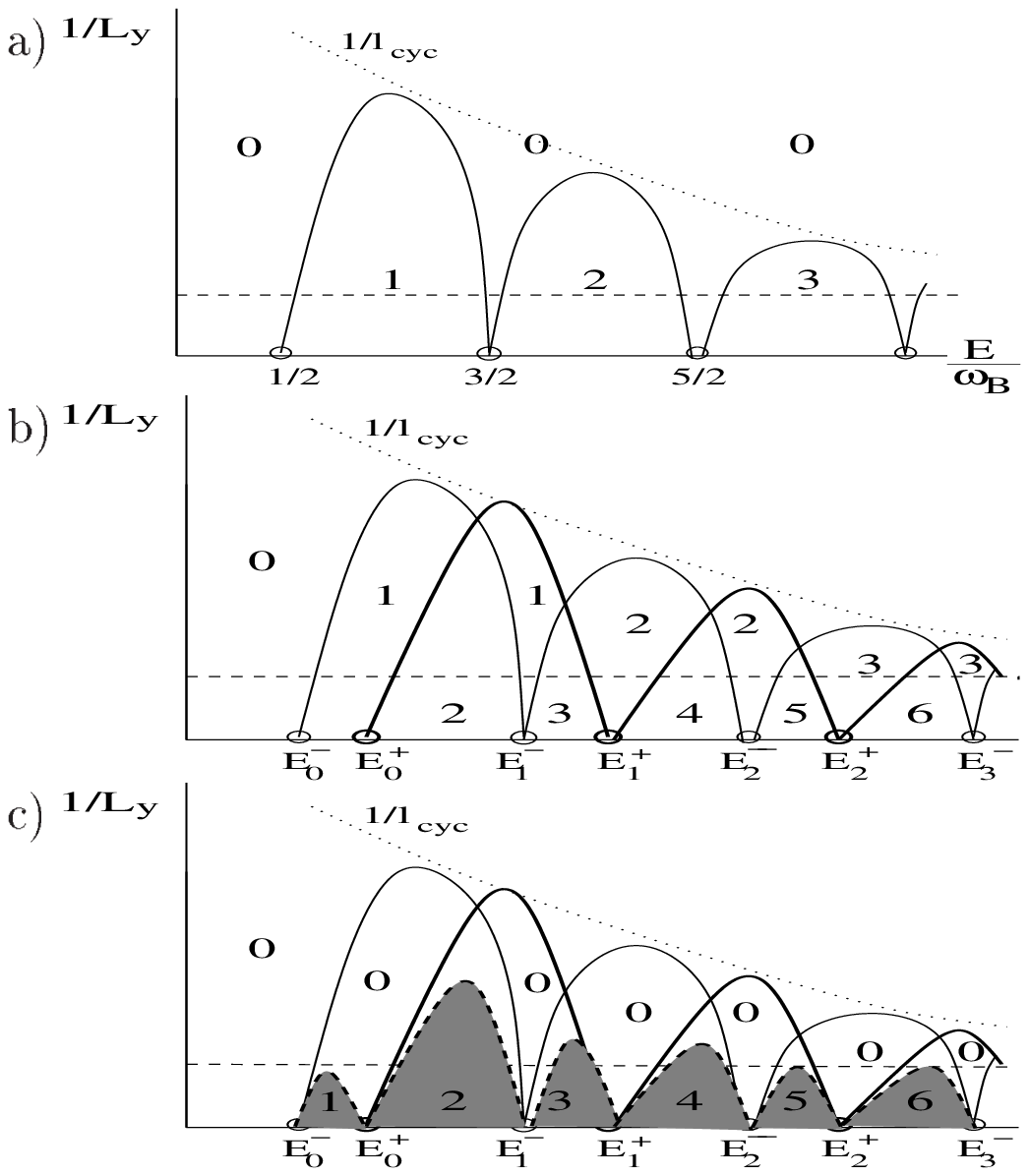} 
\vspace{-10.4cm} 
\caption[]{  Schematic phase diagram of a quantum Hall wire, 
 for  $L \gg L_y$.  The 
full lines indicate    jumps between integer
 plateau values of  conductance $G$,  in units of $e^2/h$,
as denoted by integers. 
(a) Spinless electrons (b) Electrons with spin and  Zeeman--Splitting 
 $(E_n^+ - E_n^-)/\hbar \omega_B = g_Z/2 $ without spin flip, and 
(c) with strong spin flip rate.
 Dotted line:  inverse cyclotron length, $1/l_{cyc}$. 
 For a particular value of inverse width $1/L_y$ (dashed line) 
 a sequence of conductance plateaus  similar to Fig. 1 is obtained. 
 In the limit of $1/L_y \rightarrow 0$ there are  delocalisation critical points
  of bulk states (circles). 
 }
 \label{phase} 
\vspace{-.5cm} 
\end{center} 
\end{figure} 
 In the presence of long range interactions,  interacting edge states  
 form a correlated Luttinger liquid. Renormalisation due to edge plasmon 
 excitations   enhances
 the interedge scattering amplitude\cite{kane}.
 This results in  a decrease of  the localisation length of the edge states. 
 Accordingly, the width of the plateaus of lower Landau bands is ecpected to be reduced
 due to the Luttinger liquid correlations.
 Similarly, in the fractional quantum Hall regime, 
 where the edge state excitations are strongly correlated even
 without long range interactions\cite{kane}, 
 a reduction of the localisation length of edge states is expetced as function of the 
 filling factor $\nu$. 

 In 3--D layered systems in 
 perpendicular  magnetic field,  surface states  form 
 2D chiral metals in  plateau regions where  bulk states are localised\cite{chalker}. 
 There
are  transitions between these
 2D chiral metals and  insulating
  states in long quasi-1D wires of  layered electron systems. 

We conclude, that in quantum Hall bars of finite width $L_y \ll \xi_n$ at low
temperatures  quantum phase transitions occur between
 extended chiral  edge states and a quasi-1D insulator. 
These are driven by the  crossover from 2D to 1D localization of  bulk states. These
metal-insulator transitions
{\it resemble } first-order
phase transitions in the sense that the localization length
 abruptly jumps between exponentially large  and finite values.
 In the thermodynamic limit, {\it fixing the aspect ratio $c = L/L_y$, 
 when sending $L \rightarrow \infty$, then $c \rightarrow \infty$, 
  the two--terminal  conductance 
 jumps between exactly integer values and     zero}. 
 The transitions occur at energies where the localization
length of  bulk states is equal to the geometrical wire width. Then,  $m$ edge
states  mix and  electrons are free to diffuse between
the wire boundaries but become Anderson localized along the wire.
 At finite temperature, this
phenomenon can be observed, when the phase coherence length exceeds the
quasi-1D localization length in  centers of  Landau bands, $L_{\varphi}
> \xi_n$. It may accordingly be called {\it mesoscopic
 Chiral Metal--Insulator Transition}. 
 In Hall bars of large aspect ratios at low temperatures one should observe 
transitions of the two-terminal resistance from integer quantized plateaus,
$R_n = h/n e^2$ to a Mott variable-range hopping regime of exponentially
diverging resistance. Such experiments would yield new information about edge
states in quantum Hall bars.
At higher temperature, when $L_{\varphi}< \xi_n$, the conventional form of the
integer quantum Hall effect is recovered \cite{klitzing}.
   
{\bf Acknowledgments} The authors gratefully acknowledge useful discussions
with Bodo Huckestein,  Mikhail Raikh,   Isa Zarekeshev, and  
Grzegorz Michalek.
 This research was
supported by  German Research Council (DFG), Grant No. Kr 627/10,  
 Schwerpunkt "Quanten-Hall-Effekt",
 and  by EU TMR-network Grant. No.  HPRN-CT2000-0144.

\end{document}